\documentclass[12pt]{iopart}
\usepackage{graphicx}
\usepackage{iopams}

\newcommand{\s}{~}

\graphicspath{{./figures/}}

\begin{document}

\title[]{ Neutron diffraction study of stability and phase transitions in Cu-Sn-In alloys as alternative Pb-free solders}
\author{ G Aurelio$^1$, S A Sommadossi$^2$, G J Cuello$^3$}
\address{$^1$ Consejo Nacional de Investigaciones Cient\'{\i }ficas y
T\'{e}cnicas, Centro At\'{o}mico Bariloche - Comisi\'{o}n Nacional
de Energ\'{\i}a At\'{o}mica, Av. Bustillo 9500, 8400 S. C. de
Bariloche, RN, Argentina}
\address{$^2$ IDEPA-Consejo Nacional de Investigaciones Cient\'{\i }ficas y
T\'{e}cnicas --Facultad de Ingenier\'ia, Universidad Nacional del Comahue, Buenos Aires 1400, 8300 Neuqu\'en, Argentina.}

\address{$^3$ Institut Laue Langevin, F-38042 Grenoble, France}
\ead{gaurelio@cab.cnea.gov.ar}
\date{\today}

\begin{abstract}

In this work we present an experimental study of structure and phase stability in ternary Cu-Sn-In alloys around 55 at.\% Cu in the temperature range 100$^{\circ}$C $\leq T \leq$ 550$^{\circ}$C. We have followed in real-time the sequence of phase transformations in succesive heating and cooling ramps, using state-of-the-art neutron powder thermodiffractometry. These experiments were complemented with calorimetric studies of the same alloys. Our results give experimental support to the current assessment of the ternary phase diagram in this composition and temperature range, yielding the sequence of transitions $\eta \rightarrow (\eta + L) \rightarrow (\varepsilon + L)$ with transformation temperatures of 210$^{\circ}$C and 445$^{\circ}$C, respectively. The use of neutrons allowed to overcome common difficulties in phase identification with powder XRD due to absorption and preferred orientation issues. Even the transitions to liquid phases could be successfully identified and monitored \textit{in situ}, turning this technique into a valuable tool for phase diagram studies of emerging lead-free solder candidates.  
\end{abstract}


\maketitle 
%%%%%%%%%%%%%%%%%%%%%%%%%%%%%%%%%%%%%%%%%%%%%%%%%%%%%%%%%%%%%%%%%%%%%%%%%%%%%%%%%%%%

\section{Introduction \label{Introduction}}

Sn-based alloys have gained incresing attention in the last years, especially since the implementation of regulations to eliminate Pb from the electronic industry. Tin, the most used Pb replacement, forms intermetallic phases (IPs) with most conductor metals. These IPs usually form a layer between the metallic substrate and the solder alloy, which is indicative of a good metallurgical bonding. Despite their essential presence in the bonds, IPs must be carefully monitored, because they are often brittle and may constitute a source of structural defects which may affect the performance of the devices during service. For this reason, the formation and evolution of IPs in solder alloys is an issue of fundamental interest\s\cite{08Din}. It has been pointed out \cite{05Lau} that systematic investigations are still lacking related to the formation of IPs and especially to the effect of alloying elements on their properties in systems such as Cu-Sn, Ni-Sn, Au-Sn, Ag-Sn and In-Sn--the binary system with the lowest eutectic point at only 120$^{\circ}$C.

Therefore, we discuss in the present paper alloys from the Cu-Sn-In ternary system, which has gained attention as a potential solder for alternative methods such as transient-liquid-phase-bonding, and for high-performance and/or high-temperature applications. The equilibrium phase diagram of Cu-Sn-In is still under evaluation \cite{07Vel} and only a few systematic experimental studies on the ternary system have been conducted \cite{72Koe,08Lin,09Lin}. Of particular interest for industry and bonding technologies is the IP called $\eta$-phase, which is a stable phase both in Cu-In (Cu$_2$In) and in Cu-Sn (Cu$_6$Sn$_5$) alloys, and is the most common IP found in the interface Cu-solder. The $\eta$-phase Cu$_6$Sn$_5$ has two structural forms: the high--temperature hexagonal form (NiAs-type) called in the following HT-$\eta$ and the low temperature, long period superstructure with monoclinic symmetry called LT-$\eta$, stable at room temperature and below 186$^{\circ}$C in the binary Cu-Sn system\s\cite{90Mas}. During the soldering process, however, the HT form is usually metastably retained, although during subsequent use of devices the HT$\rightarrow$LT transformation may occur. Such transition is accompanied by a specific volume change of about 2.15\%  which is highly nocive for the performance of soldered joints as it produces cracks \cite{10Nog,10Sch}. Defects and impurities may also play a significant role in the stability and kinetics of these phase transformations. To sum up, to make reliable bonds for high-performance technological applications, it turns essential to understand the crystallography, structural properties, stability and transformation mechanisms of the $\eta$-phase and other IPs of the Cu-Sn-In system. 
The present work focusses on the thermal stabilty and phase transitions of the less-studied ternary $\eta$-phase in Cu-rich Cu-Sn-In alloys. We have followed the crystal structure using \textit{in situ} thermodiffraction experiments using neutrons, a technique that proved to have major advantages over X-ray diffraction for this kind of metallic systems \cite{03Aur,05Aur-a,06Mar,12Aur-a}. Phase transitions have also been monitored by differential scanning calorimetry (DSC) measurements. The main objective of the present work is to follow the stability of the $\eta$-phase in the temperature range 25$^{\circ}$C$<T<$450$^{\circ}$C for three selected ternary samples.

%%%%%%%
\section{Experimental methods}

The samples discussed in this work correspond to samples labelled S4, S5 and S6 in Ref.\s\cite{12Aur-a}, and its compositions are listed in Table\s\ref{t:1}. The three alloys were annealed at $300^{\circ}$C for 3 weeks to promote homogenization, and the ingots were then rapidly quenched to $0^{\circ}$C. Powders were obtained by manually grinding the ingots for 10 minutes in an agate mortar.

Neutron thermodiffraction experiments were performed in the D1B two-axes diffractometer at the Institut Laue-Langevin (ILL) in Grenoble, France. The neutron spectra were collected placing the $^{3}$He multidetector of 400 cells in a cylindrical geometry centered at the sample. We used a wavelength of $2.52$ \AA, an angular span of $80{^{\circ }}$ and steps of 0.2${^{\circ }}$. Silicium and alumina standards were used to calibrate the wavelength, yielding the value $\lambda=2.5275$ \AA. Samples measured at high--temperature were placed in vanadium--foil sample holders, whereas room--temperature measurements were performed using conventional thick vanadium cylinders of 8mm diameter. In all cases the sample holders were almost completely filled with powder. The neutron beam size on the samples was 14mm$\times$50mm. Measurements were performed inside a standard vanadium furnace under a vacuum of $10^{-4}$ mbar. To accelerate the cooling process, Ar was injected into the furnace between $250{^{\circ }}$C and $100{^{\circ}}$C before opening the furnace to prevent oxidation. 

The neutron flux on the sample was of about 10$^{6}$ n/(s cm$^{2})$, which allowed us to follow the evolution of the samples on warming by collecting diffraction patterns every 2 minutes between room temperature and  $450{^{\circ}}$C with intermediate 30-minute dwells at  $300{^{\circ}}$C. A heating rate of 1.6$^{\circ}$/min was applied. In Table \ref{t:1} we summarise the heat treatments applied to the samples. During the temperature dwells, 5-minute patterns where added up to enhance the
statistics. The diffraction data were qualitatively analysed and plotted using the \begin{scriptsize} LAMP \end{scriptsize} software \cite{LAMP} and then processed with \begin{scriptsize} FULLPROF\end{scriptsize}, using models from the higher resolution diffraction data collected at the D2B diffractometer and reported in Ref.\s\cite{12Aur-a}. Additional bulk material from the same ingots was manually ground to obtain a powder suitable for differential scanning calorimetry (DSC). Calorimetric curves were collected in Al crucibles using a DSC-2910 TA Instrument under an +Ar flow of 120 ml/min. Successive warming and cooling ramps were performed with rates between $2$ and $25{^{\circ}}/$min.

\begin{table*}
 \caption{List of Cu-In-Sn alloys studied in the present work. The nominal composition and the applied thermal treatments are indicated.}
 \begin{center}
\begin{small}
\begin{tabular}{lcccl}
Sample & \multicolumn{3}{c}{Nominal at.\%} &  Thermal treatment \\ 
 & Cu & In & Sn &  \\ \hline \noalign{\smallskip}
&    &   & &   3 weeks @ $300{^{\circ}}$C - quenched in iced water\\ 
S4 & 55 & 5 & 40    & warming to $450{^{\circ}}$C @ 1.6 $^{\circ}$/min - 30 min dwell @ 450$^{\circ}$C \\\cline{1-4}
&    &   &     & cooling to $100{^{\circ}}$C @ 8 $^{\circ}$/min \\
S5 & 58 & 12  & 30     & warming to $300{^{\circ}}$C @ 1.6 $^{\circ}$/min - 30 min dwell @ 300$^{\circ}$C \\\cline{1-4}
&    &   &     & warming to $450{^{\circ}}$C @ 1.6 $^{\circ}$/min - 30 min dwell @ 450$^{\circ}$C \\
S6 & 60 & 20  & 20    & cooling to room temperature \\ \hline
\end{tabular}
\end{small}
 \end{center}
\label{t:1}
\end{table*}

\section{Results  \label{s:Results}}

\subsection{As-quenched state}

The three selected samples for this work had been previously studied using high-resolution ND at room temperature, in their as--quenched state, revealing that they are mostly constituted by the HT-$\eta$ phase\s\cite{12Aur-a}. However, we mentioned there that the alloys showed certain indications of some LT-$\eta$. This second phase was not included in those refinements for several reasons, most importantly because the phase fraction seems small, and the weak reflections which distinguish the LT-$\eta$ from the HT-$\eta$ are very broad and unresolved, suggesting a small crystallite size. Although this omission was not determining in the systematization of the structural parameters of the primary HT-$\eta$ phase, it now turns important when analysing the phase transitions and equilibrium conditions in our thermodiffraction experiments, and helps to understand the calorimetric data.

\begin{figure}[tb]
\centering \vspace{3mm}
\includegraphics[width=0.8\linewidth]{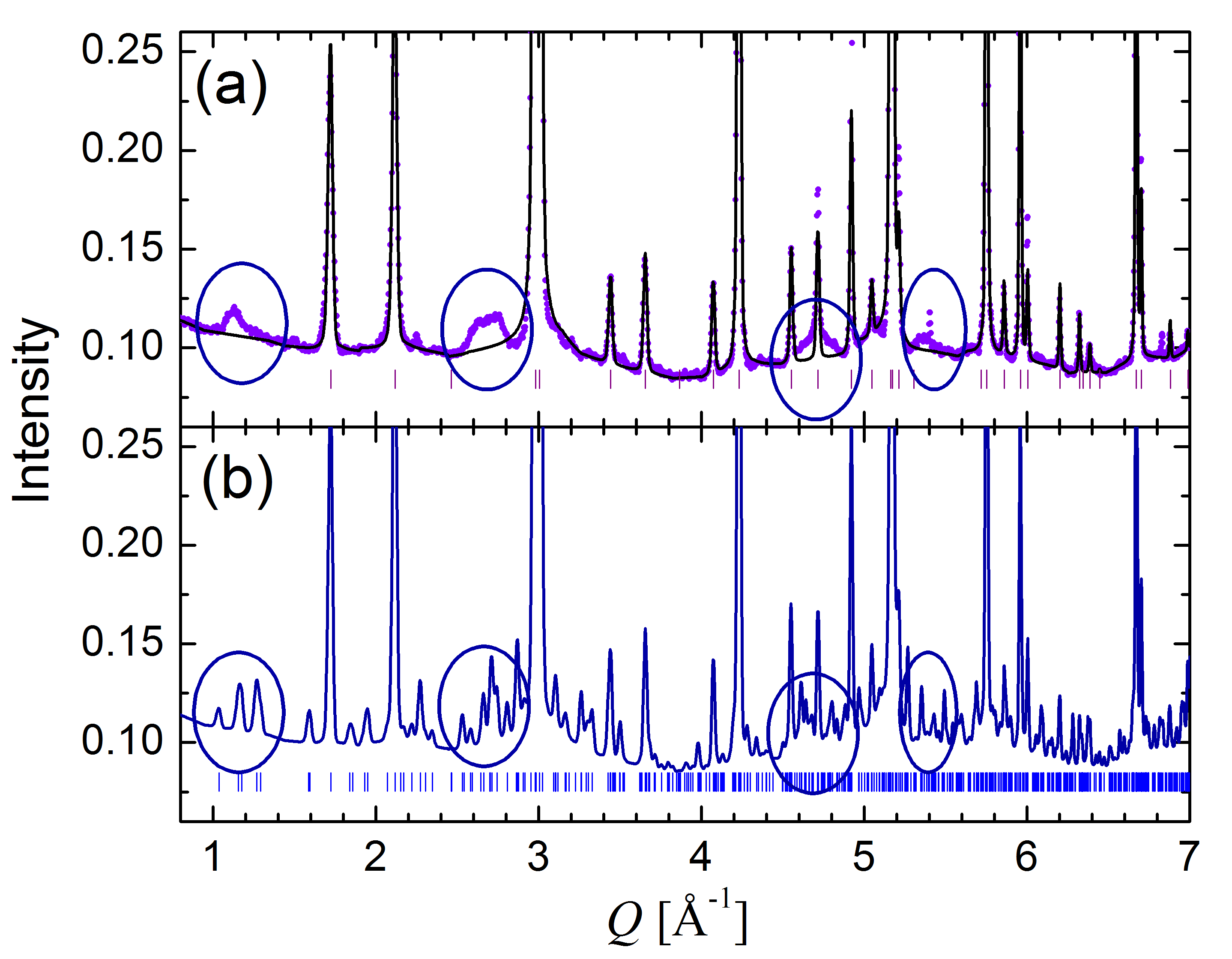}
\caption{(a) Rietveld refinement for sample S4 with nominal 55 at.\% Cu- 40 at.\% Sn- 5 at.\% In queched from $300{^{\circ}}$C, from data collected at room temperature in D2B using $\lambda=1.593$\s\AA \cite{12Aur-a}. The vertical bars at the bottom indicate the HT-$\eta$ (hexagonal) Bragg reflections. Major discrepancies between the single-phase model and the D2B data are highlighted with circles. (b) A theoretical diffractogram calculated with the available crystallographic information on LT-$\eta$ (monoclinic $C2/c$, ICSD 106530) and the same instrumental resolution file from D2B. The circles highlight the same areas as in (a).} \label{f:S4-ref}
\end{figure}

In Fig.\s\ref{f:S4-ref}(a) we recall a magnification of the Rietveld refinement of sample S4\s\cite{12Aur-a}, with nominal 5 at.\% In and 40 at.\% Sn, measured at room temperature with high-resolution ND. The solid line represents a fit using only the HT-$\eta$ phase with In occupying the $2c$ site in the adequate nominal proportion. It is clear in this magnification that low-intesity data are not completely accounted for using just the HT-$\eta$ phase (Bragg reflections indicated by vertical bars at the bottom), and neither by the $\varepsilon$-phase, present in some other alloys in Ref.\s\cite{12Aur-a}. We have highlighted with circles certain regions, to indicate that the single-phase model fails to account for some broad and diffuse reflections. In panel (b) we present a calculated diffractogram corresponding to the monoclinic LT-$\eta$-phase Cu$_6$Sn$_5$ (ICSD card 106530\s\cite{94Lar}). The diffractogram was modeled in similar experimental conditions, but without modifications in lattice parameters or site occupations respect to the binary phase. We can see in Fig.\s\ref{f:S4-ref}(b) that the main reflections are common to the two $\eta$ phases, and the regions highlighted in (a) are compatible with bunches of weak reflections of LT-$\eta$ (which in Fig.\s\ref{f:S4-ref}(b) have been modeled only with the instrumental resolution width).

\subsection{Phase transformations during heating}

Although the S4 sample has already 5 at\% In, it is simpler to analyse its transformations in terms of the binary Cu-Sn phase diagram \cite{90Mas}. In the region of interest, the actual transition temperatures in the ternary may be slightly shifted but the overall picture remains very similar. In fact, the proposed ternary assessment which most closely resembles the current experimental conditions is the 4 at.\% In cut between 400$^{\circ}$C and 800$^{\circ}$C compiled by the MSIT\s\cite{07Vel}, which for a 56 at.\% Cu proposes the sequence $(\varepsilon + \eta) \rightarrow (L+\varepsilon + \eta) \rightarrow (\varepsilon + L) \rightarrow (L)$ with succesive transition temperatures of 425$^{\circ}$C, 450$^{\circ}$C and 610$^{\circ}$C. Such sequence is not unexpected from considerations based on the binaries Cu-Sn and Cu-In equilibrium diagrams, beyond certain discrepancies.

\begin{figure}[tb]
\centering \vspace{3mm}
\includegraphics[width=0.5\linewidth]{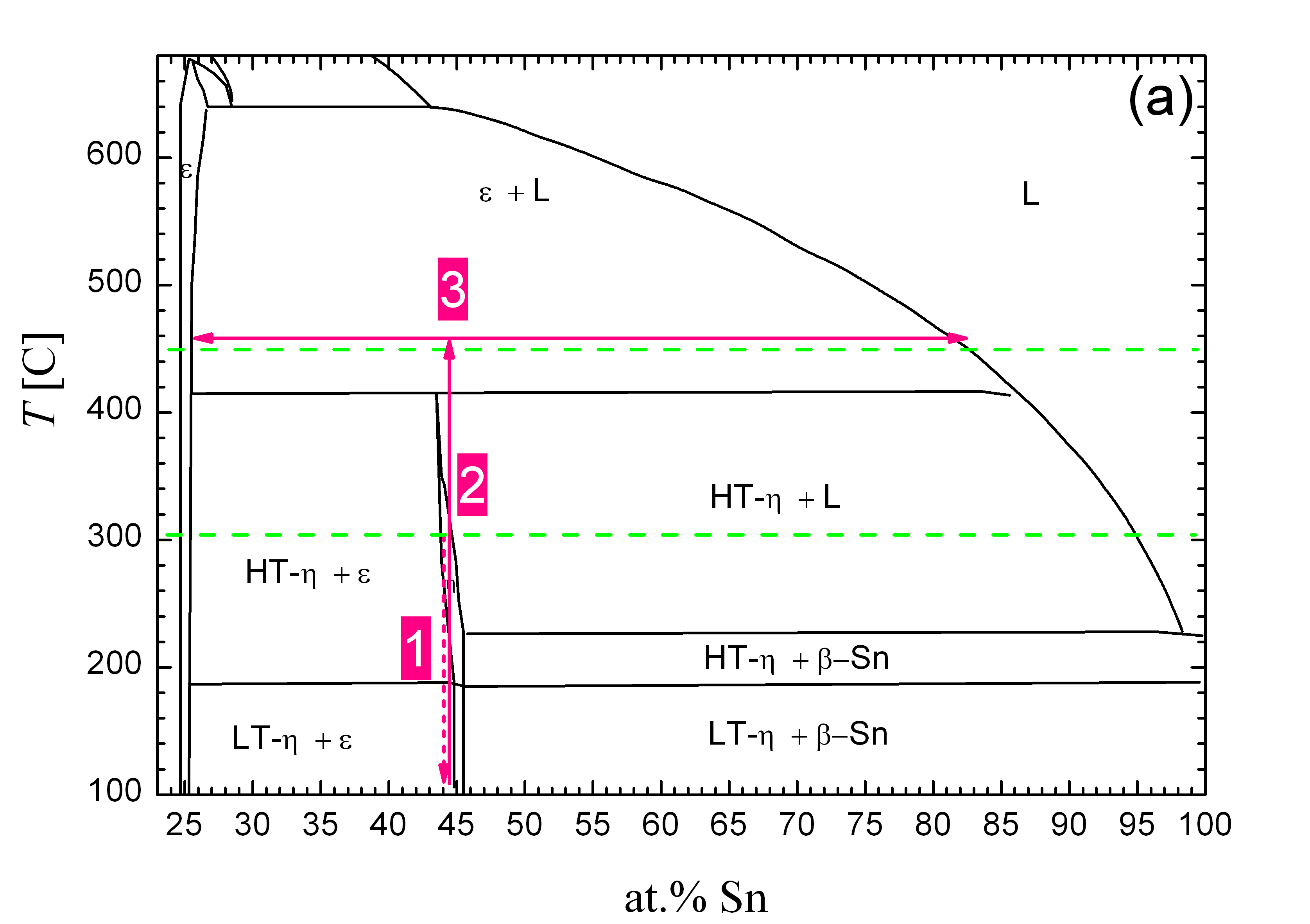}\includegraphics[width=0.5\linewidth]{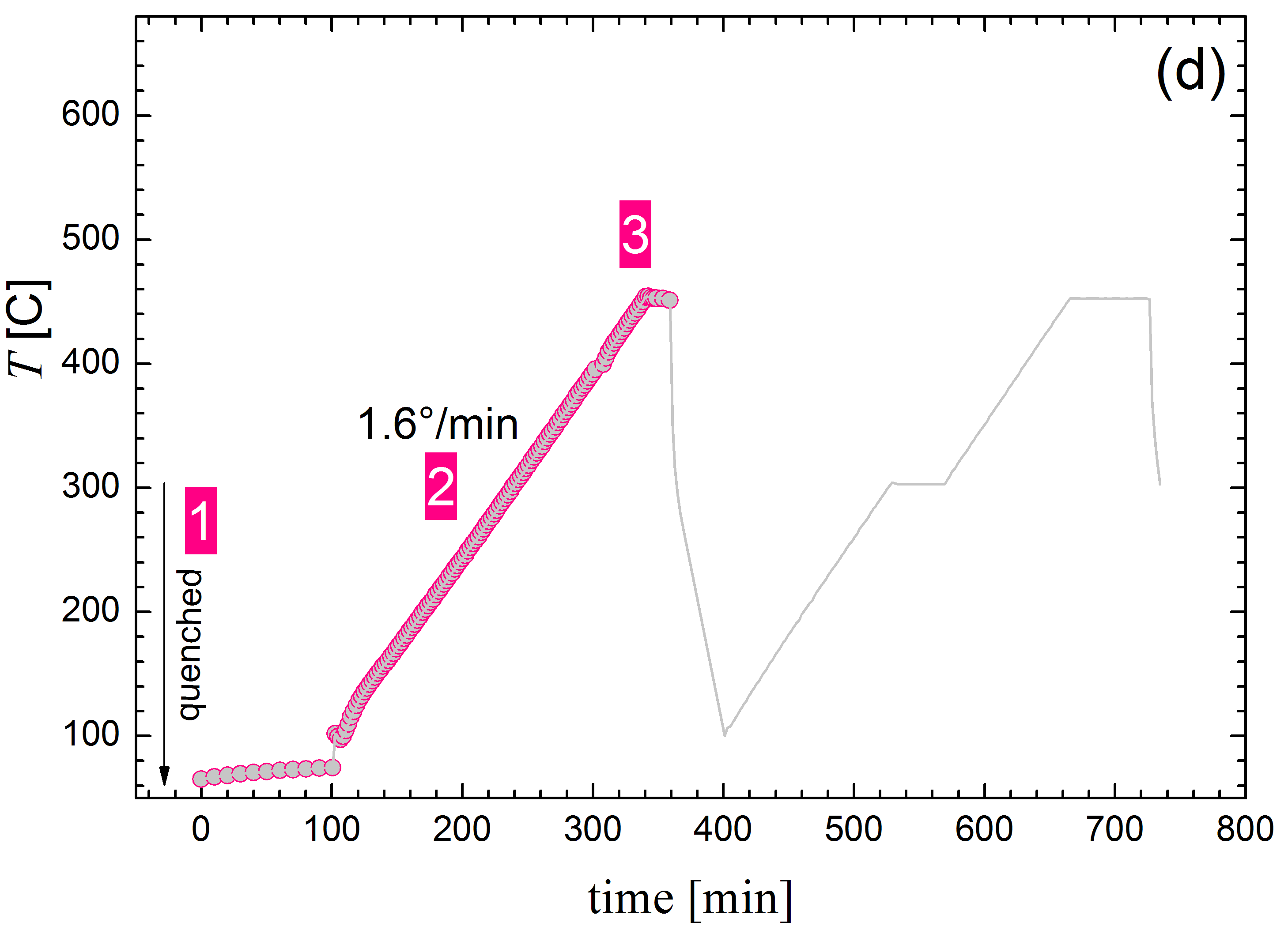}
\includegraphics[width=0.5\linewidth]{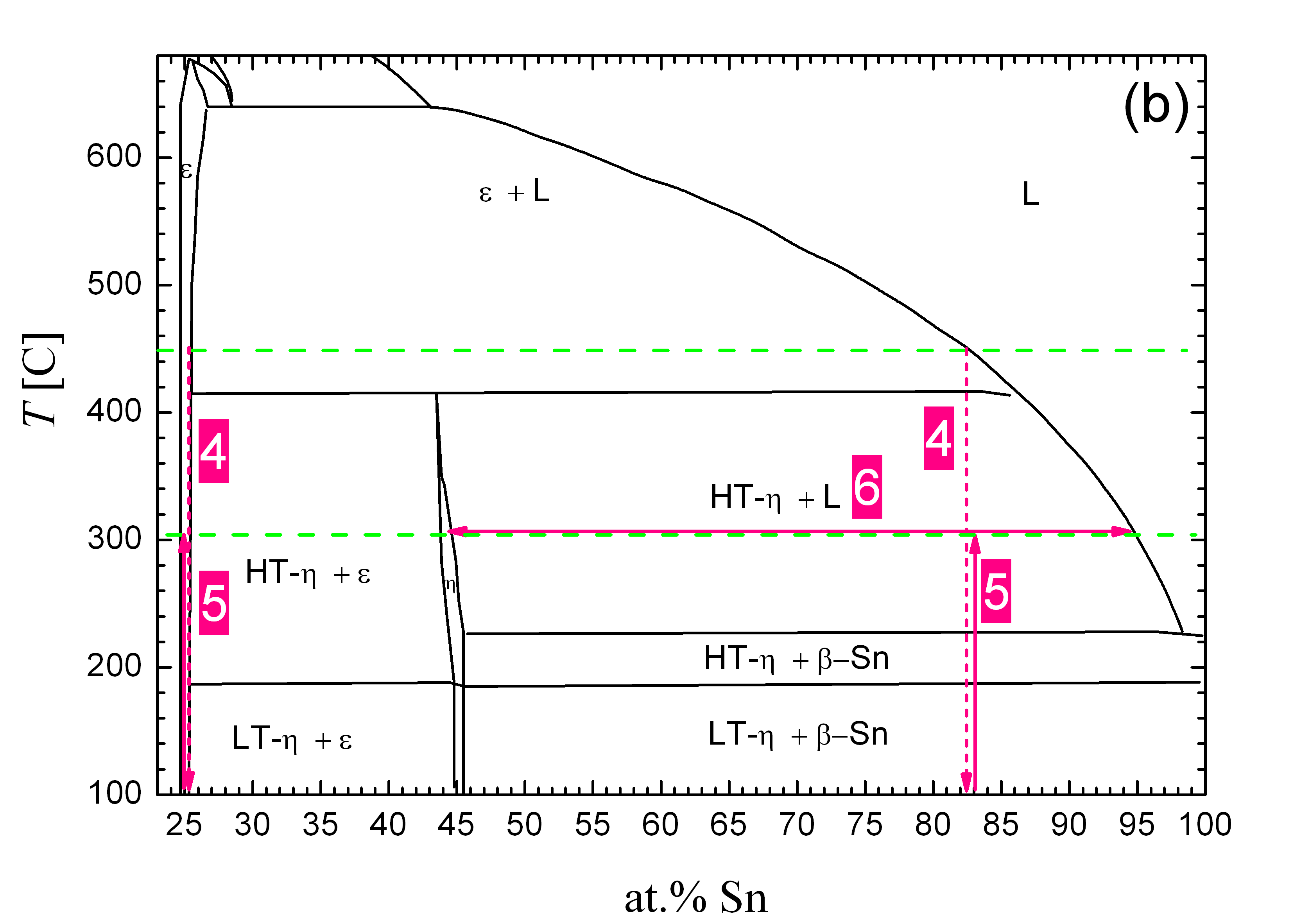}\includegraphics[width=0.5\linewidth]{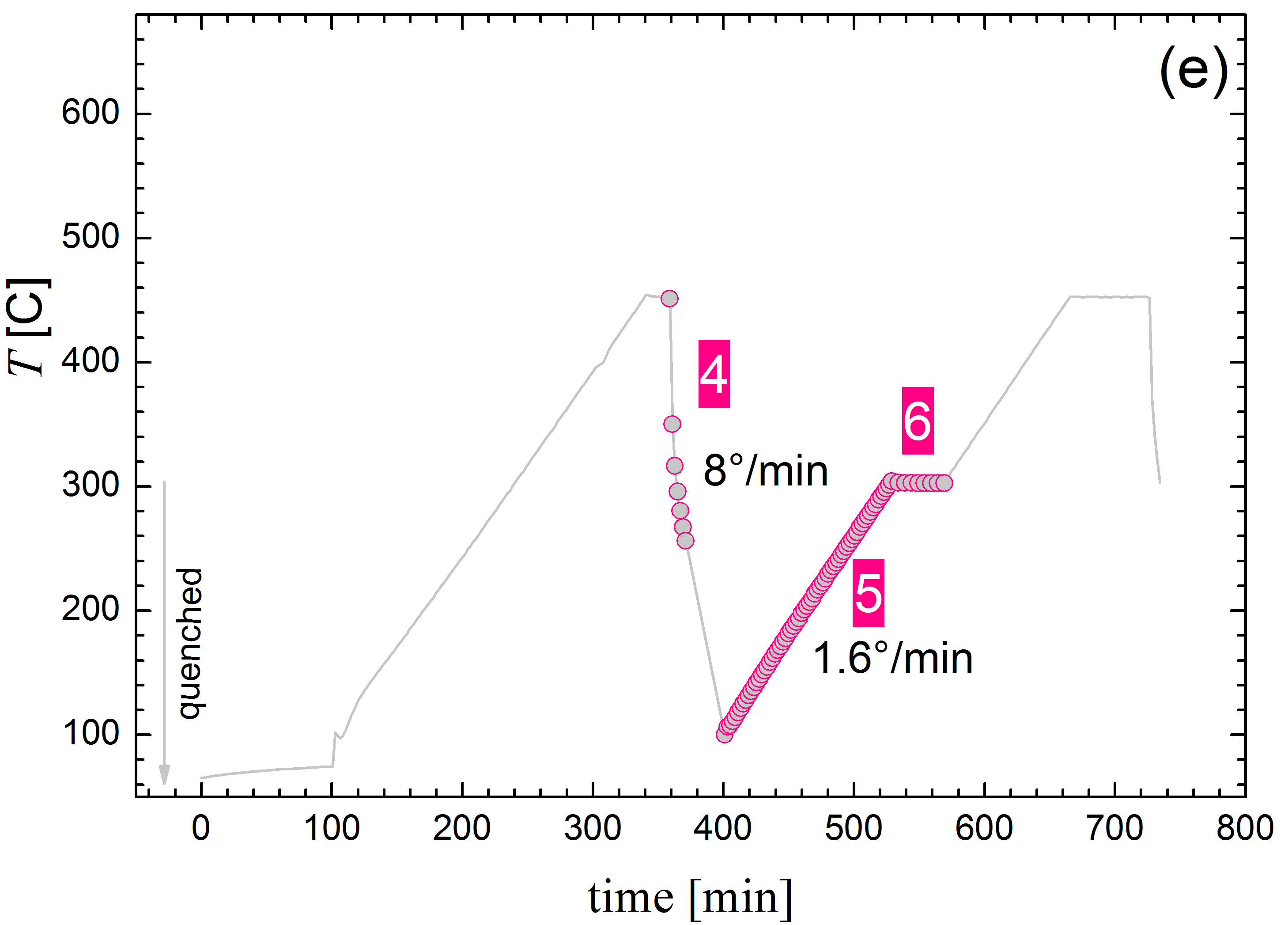}
\includegraphics[width=0.5\linewidth]{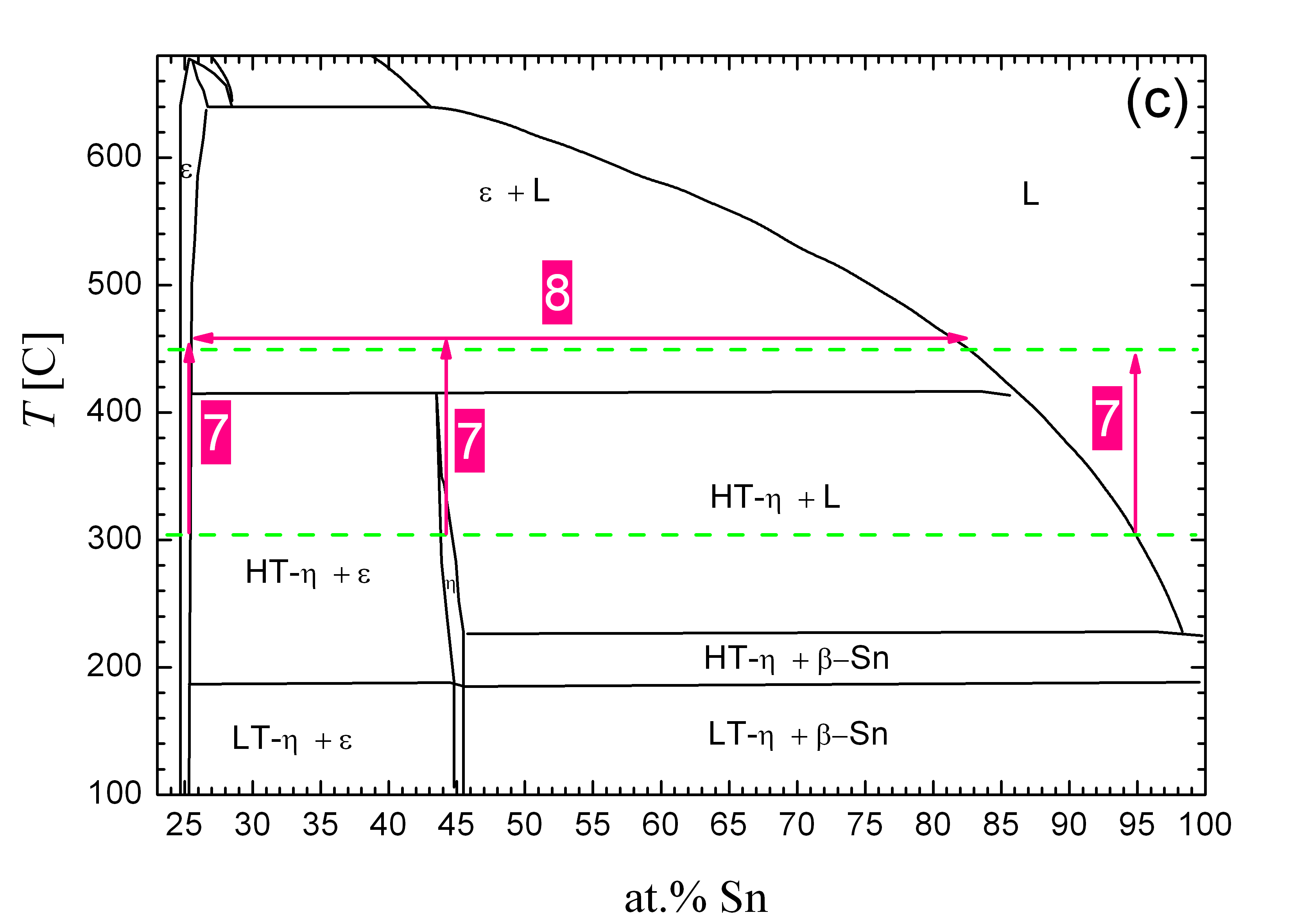}\includegraphics[width=0.5\linewidth]{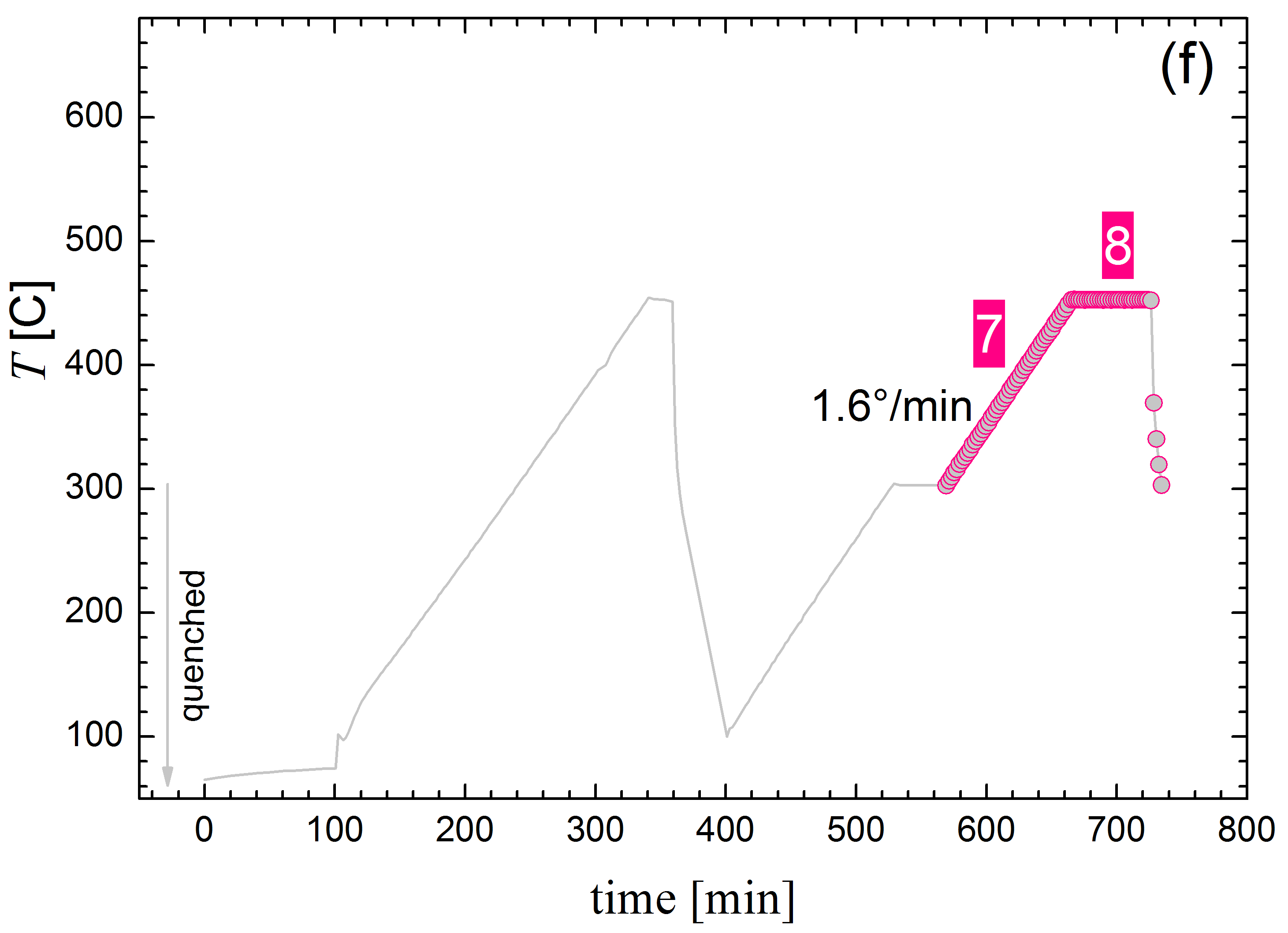}
\caption{(a,b,c) Binary Cu-Sn phase diagram after Massalski \cite{90Mas}. The numbered arrows denote the thermal history of the sample during the ND experiments, which correspond to the temperature ramps in the right-hand graphs (d,e,f). The two temperature dwells at 300$^{\circ}$C and 550$^{\circ}$C are indicated by dashed lines. Dotted arrows indicate quenching or rapid cooling, not allowing for equilibrium conditions to be attained.}  \label{f:Cu-Sn}
\end{figure}

\begin{figure}[tb]
\centering \vspace{3mm}
\includegraphics[width=0.8\linewidth]{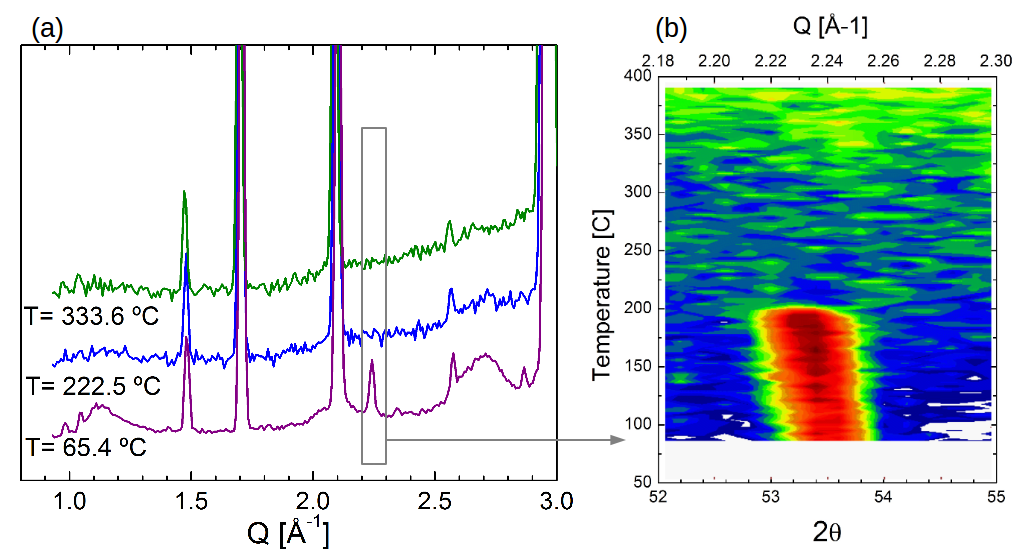}
\caption{ (a) Magnification of the low-$Q$ region for sample S4 diffractograms collected at different temperatures in instrument D1B using $\lambda=2.52$\s\AA. (b) Projection to the $2 \theta - T$ plane of the thermodiffractograms showing a Bragg reflection (possibly (1 0 1) from $\beta-$Sn), which vanishes between 200$^{\circ}$C and 220$^{\circ}$C. Data were collected on warming with $\lambda=2.52$\s{\AA} between 65$^{\circ}$C and 400$^{\circ}$C.} \label{f:S4-ramp1}
\end{figure}

The initial state for our S4 sample is the metastable retention at room temperature of the stable phase at 300$^{\circ}$C. This process is indicated in Fig.\s\ref{f:Cu-Sn}(a) with an arrow labelled 1. In Fig.\s\ref{f:Cu-Sn}(b) the thermal history is sketched, highlighting the relevant stages discussed below. The following stage (2) corresponds to a heating ramp between room temperature and 450$^{\circ}$C. When sample S4 was measured at the D1B diffractometer, and heated \textit{in situ}, the diffractograms showed subtle changes around  $200{^{\circ}}$C. In Fig.\s\ref{f:S4-ramp1}(a) we present some selected diffractograms in the low-$Q$ region collected at different temperatures for two minutes. We have used a $Q$ scale in Fig.\s\ref{f:S4-ref} and Fig.\s\ref{f:S4-ramp1}(a) to allow for comparison of diffractograms collected at different wavelengths. At the lowest temperature measured, 65$^{\circ}$C, the D1B data in Fig.\s\ref{f:S4-ramp1}(a) coincides with room temperature D2B data (Fig.\s\ref{f:S4-ref}(a)) except for the peaks at 1.48\s\AA$^{-1}$ and 2.6\s\AA$^{-1}$ which probably come from the furnace. However, as soon as the sample is heated above 220$^{\circ}$C, a set of peaks vanishes, in particular the wide bumps at 1.1\s\AA$^{-1}$ and 2.7\s\AA$^{-1}$ which can be associated to the LT-$\eta$ as discussed in the previous subsection. Nogita \textit{et al.}\s\cite{09Nog} also observed using synchrotron XRD the coexistence of hexagonal (HT) and monoclinic (LT)-$\eta$ phase in Cu$_6$Sn$_5$ alloys quenched from 400$^{\circ}$C, the monoclinic phase only being noticed by very weak additional reflections.

\begin{figure}[tb]
\centering \vspace{3mm}
\includegraphics[width=0.6\linewidth]{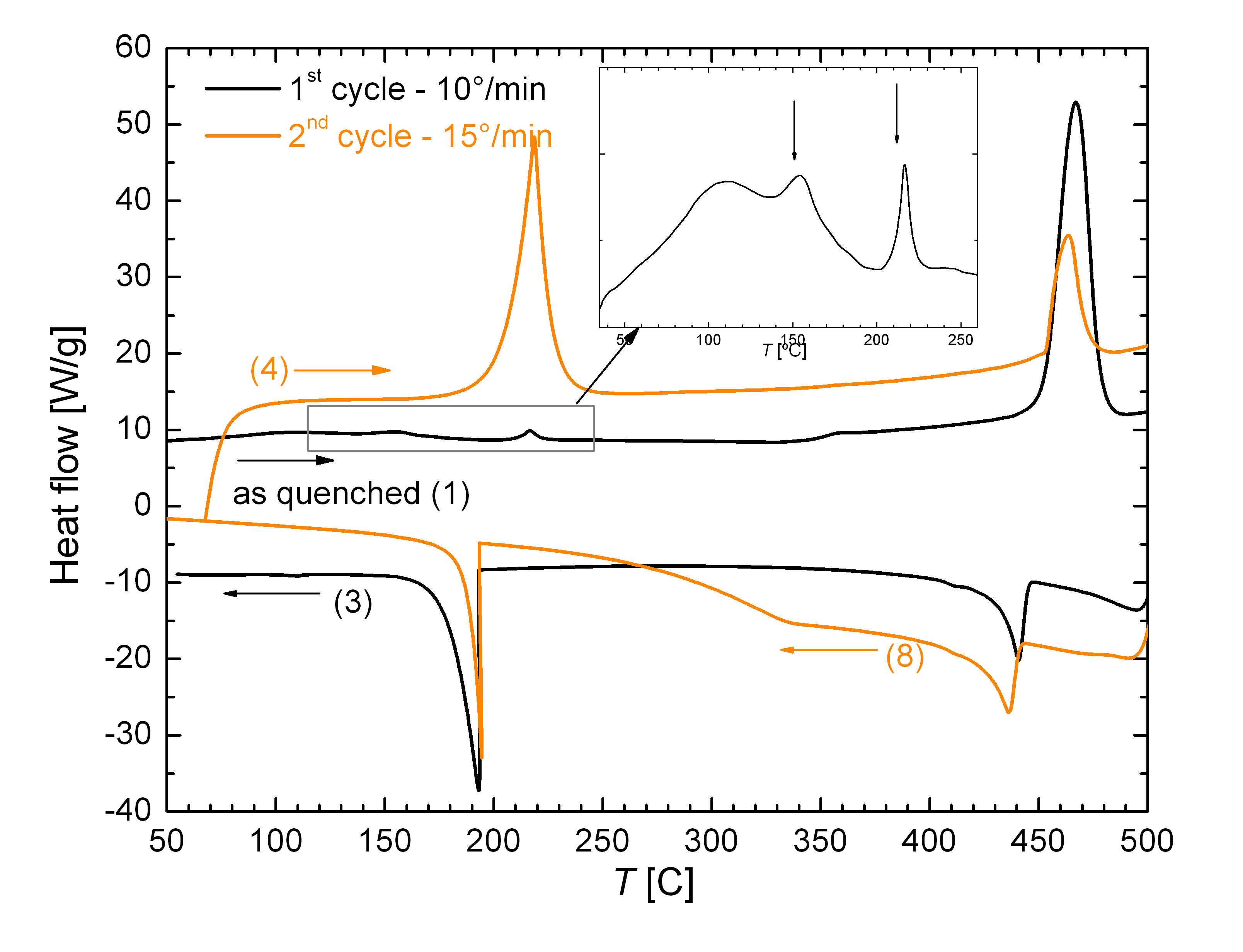}
\caption{DSC curves collected on warming for
sample S4. The first cycle (dark curve) corresponds to the as-
quenched sample, while the second curve (light curve) was collected immediately after the fist cycle and corresponds to another thermodynamical state. The curves were obtained under N$_2$ flow for a heating/ cooling rate of $10^{\circ}/$min (first cycle) and $15^{\circ}/$min (second cycle). Numbers in brackets correspond to the stages represented in Fig.\s\ref{f:Cu-Sn}.} \label{f:S4-dsc}
\end{figure}

 DSC measurements of this sample also evidence small endothermic peaks at low temperature when the as-quenched sample is heated for the first time (black line in Fig.\s\ref{f:S4-dsc}). In fact, a closer inspection (inset) reveals that there are two transitions during the first heating ramp. The first peak is around 160$^{\circ}$C, while there is a second one around 216$^{\circ}$C, although both are very weak --indicating a small amount of sample involved in the transformation. Based on the results in Fig.\s\ref{f:S4-ramp1}(a), we propose that the first peak occurring at 160$^{\circ}$C corresponds to the small amount of LT-$\eta$ initially present in the as-quenched sample, transforming to the HT-$\eta$ hexagonal structure. In the literature, the LT$\rightarrow$HT transition is reported at 186$^{\circ}$C for the binary Cu$_6$Sn$_5$ \cite{72Koe} but had not yet been reported for ternary alloys. 
The second peak at 216$^{\circ}$C can be associated to the melting point of small amounts of Sn-rich seggregation at grain boundaries, as observed in the metallographies presented in \cite{12Aur-a}. In fact, the temperature evolution of the reflection at $Q=2.2$\s\AA$^{-1}$ shown in Fig.\s\ref{f:S4-ramp1}(b), can be associated to the (1 0 1) Bragg reflection of $\beta-$Sn, which has a well--established melting point of 231$^{\circ}$C. This will be reinforced by the analysis of the subsequent thermal treatments. From this analysis we conclude that between 150$^{\circ}$C and 220$^{\circ}$C weak transitions occur which result in the complete stabilisation of the metastably retained HT-$\eta$-phase.

\begin{figure}[tb]
\centering \vspace{3mm}
\includegraphics[width=\linewidth]{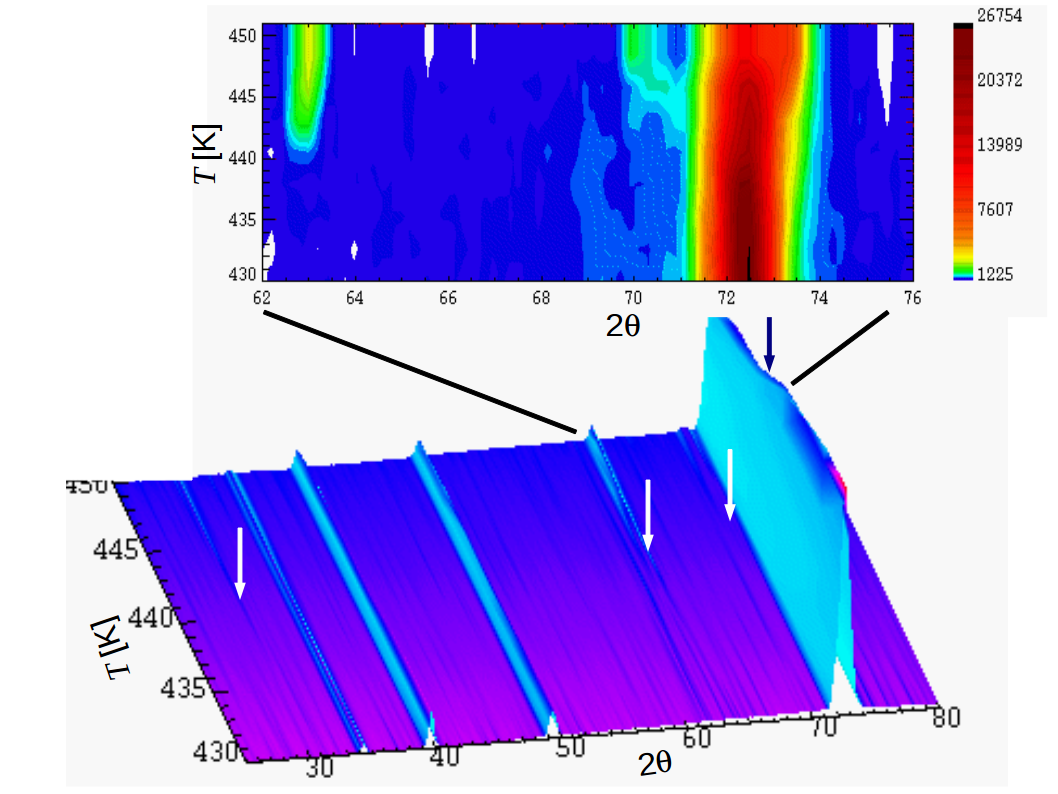}
\caption{Thermodiffractogram of sample S4 during the first heating ramp in the range 430$^{\circ}$C $< T < $450$^{\circ}$C. The gradual growth of the $\varepsilon$-phase reflections is indicated with arrows. The inset shows the projection on the $2\theta -T$ plane for a selected region highlighting the onset of $\varepsilon$ between 440 and 445$^{\circ}$C.} \label{f:rampS4}
\end{figure}

At higher temperatures, the first DSC curve shows a further transition: it corresponds to $\eta \rightarrow \varepsilon$. This transition occurs in the binary Cu-Sn system at 415$^{\circ}$C\s\cite{90Mas}, shifting to around 467$^{\circ}$C for an In addition of 5 at.\% as shown in Fig.\s\ref{f:S4-dsc}. The thermodiffractogram for the S4 sample shows that the transition occurs right at the end of the heating ramp (numbered 2 in Fig.\s\ref{f:Cu-Sn}). This is presented in Fig.\s\ref{f:rampS4}, corresponding to the final twenty degrees of ramp 2. Each diffractogram corresponds to an aquisition of 2 minutes, while the sample was being heated at 1.6$^{\circ}$/min. The arrows in Fig.\s\ref{f:rampS4} indicate the onset of the crystallographic transition $\eta \rightarrow \varepsilon$. The inset corresponds to a projection on the $2\theta-T$ plane of a selected $2\theta$ range of the diffractograms. We observe that between 440$^{\circ}$C and 445$^{\circ}$C new reflections develop which are consistent with the $\varepsilon$-phase.

\begin{figure}[tb]
\centering \vspace{3mm}
\includegraphics[width=0.5\linewidth]{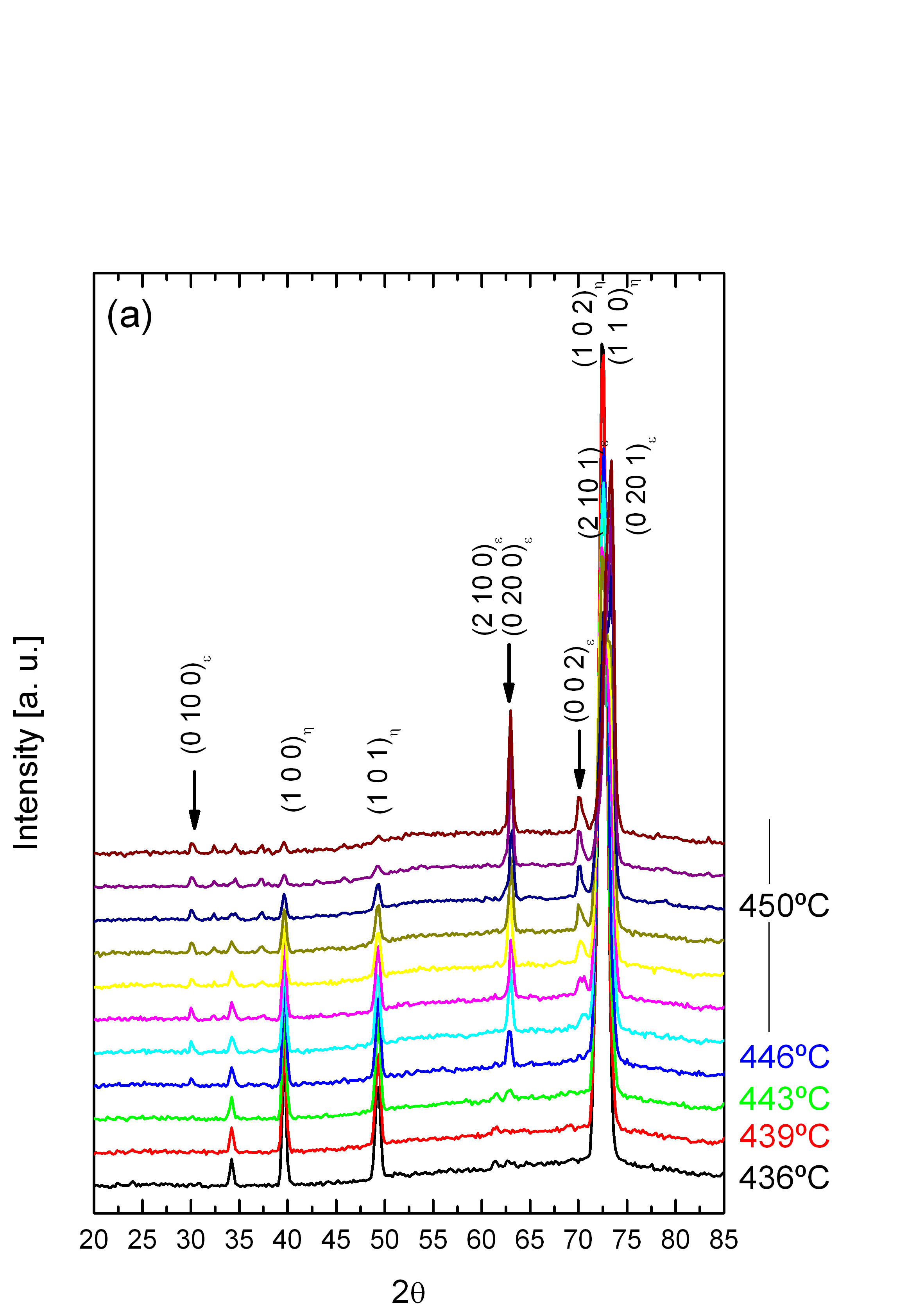}\includegraphics[width=0.5\linewidth]{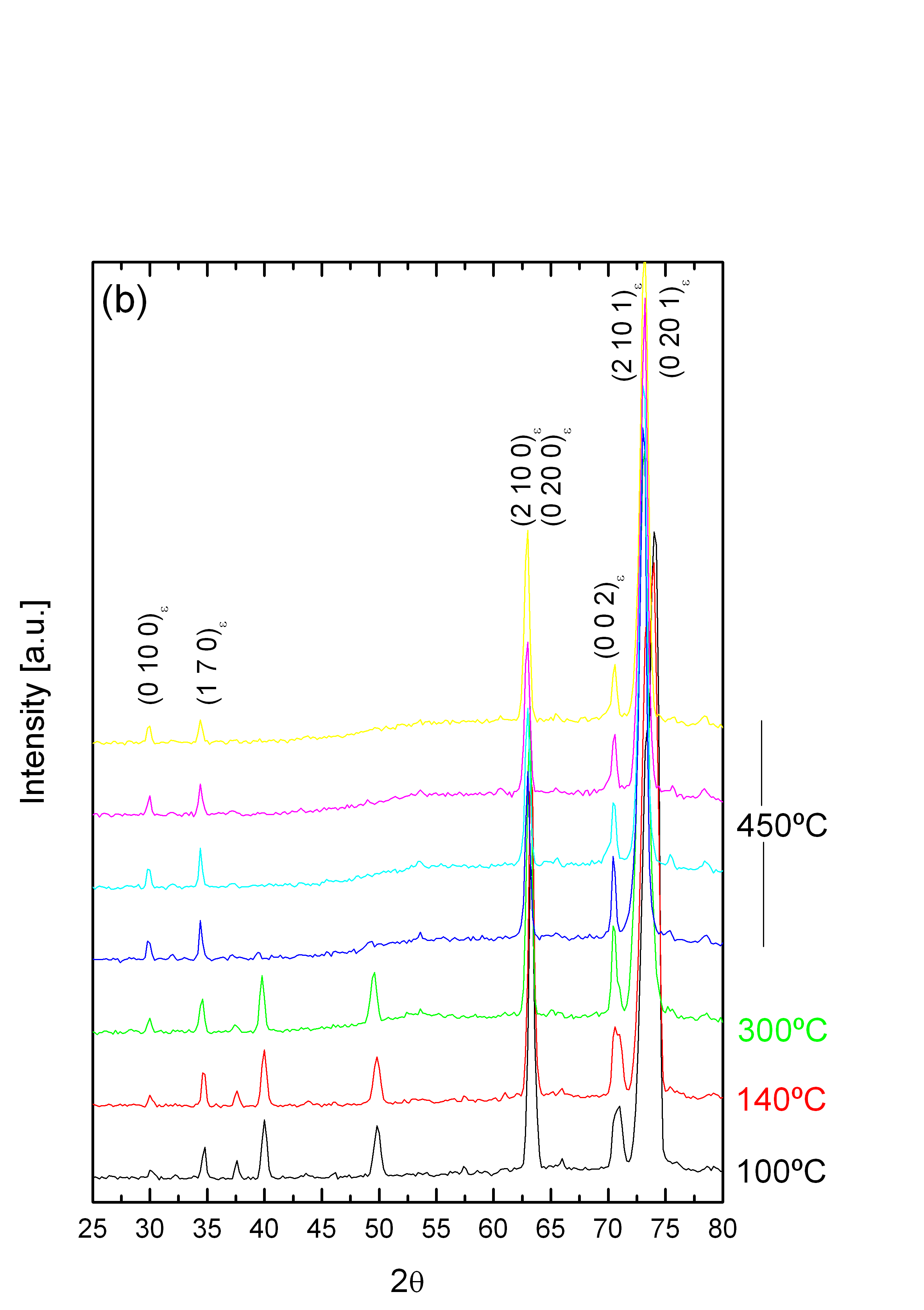}
\caption{(a) Sequence of diffractograms collected for S4 at the end of ramp (2) and during the dwell (3) at 450$^{\circ}$C, following the notation in Fig.\s\ref{f:Cu-Sn}.The reflections are indexed according to our previous work\s\cite{12Aur-a}. (b) Sequence of selected diffractograms between 100$^{\circ}$C and 450$^{\circ}$C (ramp 5-7) and during the last dwell at 450$^{\circ}$C (8). } \label{f:sequence450}
\end{figure}

According to Ref.\s\cite{07Vel}, an alloy very close in composition to S4, with 4 at.\% In and 56 at.\% Cu, should already be in a two-phase equilibrium state $\eta + \varepsilon$ at 400$^{\circ}$C, whereas at 425$^{\circ}$C a three-state equilibrium $\eta + \varepsilon + L$ is expected, and finally $\eta$ should completely transform at 450$^{\circ}$C to an $\varepsilon + L$ mixture. From the thermodiffractograms, we can only obtain information on the solid crystalline phases. What we observe there is that the sample remains in a metastable $\eta$-phase state until 440$^{\circ}$C, where $\varepsilon$ starts to nucleate, as shown in Fig.\s\ref{f:sequence450}. The evolution of the diffractograms during the dwell at 450$^{\circ}$C also show how the system is evolving towards equilibrium, gradually shifting from $\eta$ to $\varepsilon$ whereas at the same time a melting into a Sn-rich liquid is also expected from the phase diagram (Fig.\s\ref{f:Cu-Sn}(a)). We remark that the reflections marked with arrows in Fig.\s\ref{f:sequence450} for the $\varepsilon$ phase, are indexed within the orthorhombic long-period superstructure of Cu$_3$Sn (ICSD card 102103) as reported in our previous work\s\cite{12Aur-a}.

\begin{figure}[tb]
\centering \vspace{3mm}
\includegraphics[width=0.4\linewidth]{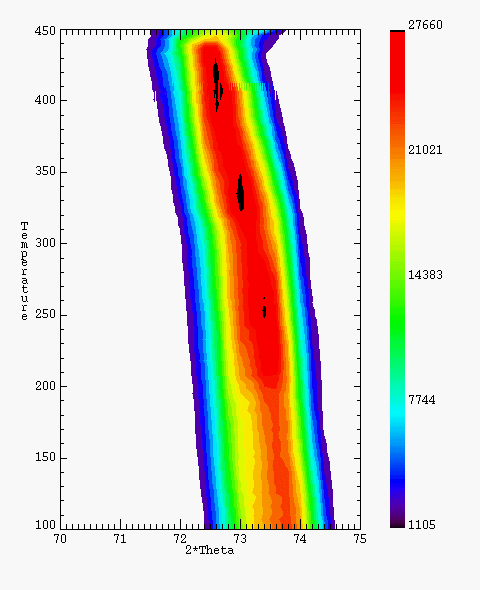}\includegraphics[width=0.4\linewidth]{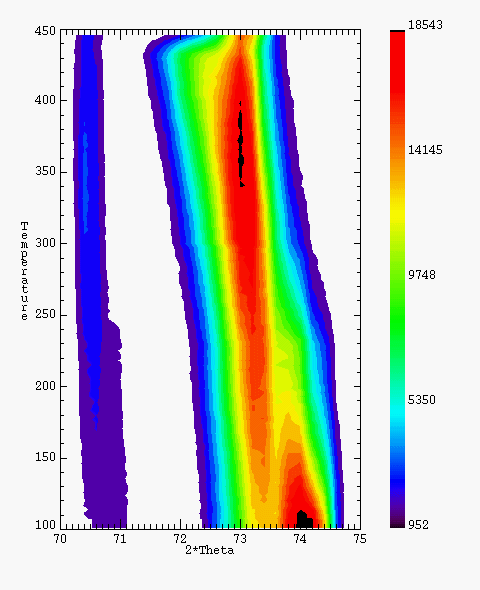}
\caption{Comparison between thermodiffractograms obtained for sample S4 during heating ramp 2 (left), and heating ramps 5-7 (right), raising from 100$^{\circ}$C to 450$^{\circ}$C at 1.6$^{\circ}$/min, as sketched in Fig.\s\ref{f:Cu-Sn}. The projection onto the $2\theta-T$ plane is presented for the most intense $\eta$ and $\varepsilon$ reflections.} \label{f:heating1vs2}
\end{figure}

\begin{figure}[tb]
\centering \vspace{3mm}
\includegraphics[width=\linewidth]{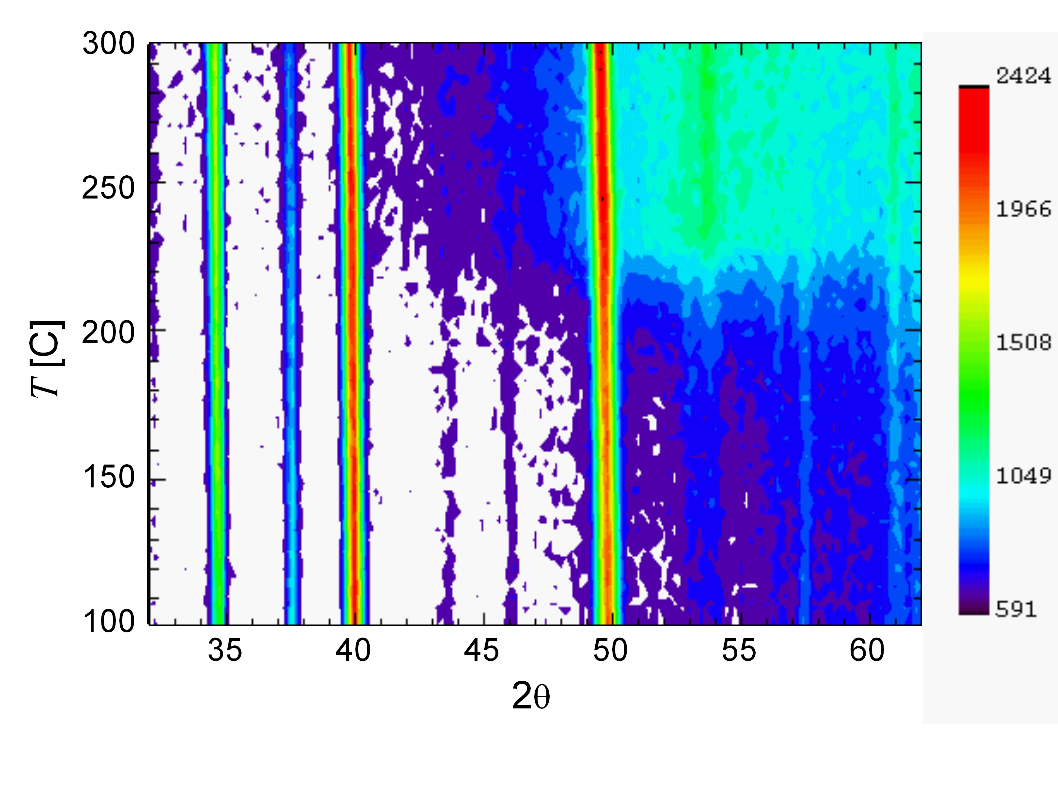}
\caption{Projection onto the $2\theta-T$ plane of the thermodiffractograms collected during ramp 5 for the S4 sample. The increase in the background intensity is adscribed to the melting into a Sn-rich liquid at ~200$^{\circ}$C.} \label{f:S4-2nd-heating}
\end{figure}

Stage 4 in the ND experiments (Fig.\s\ref{f:Cu-Sn}(e)) corresponds to a rapid cooling from 450$^{\circ}$C to room temperature, although due to the furnace inertia, the minimum temperature atteined is around 100$^{\circ}$C. The cooling rate does not allow for equilibration, almost as in a quenching process. For that reason, based on the simplified diagram in Fig.\s\ref{f:Cu-Sn}(b), we expect to find at room temperature a metastable retention of $\varepsilon +$ quenched Sn-rich liquid (probably in amorphous state). On the other hand, a reversible process would lead to the complete retransformation to the $\eta$ phase. What we actually observe is the combination of these situations. On the one hand, a certain amount of $\eta$ phase is observed along ramp 5, together with the retained $\varepsilon$ phase which was not present in ramp 2, as shown in Fig.\s\ref{f:heating1vs2}. On the other hand, the DSC experiment reveals that during the first cooling (analogue to ramp 4, but quicker), the retransformation peak $\varepsilon \rightarrow \eta$ is smaller but clearly present, whereas the Sn-solidification peak at 200$^{\circ}$C has grown considerably. All these data support the simplified image of the binary diagram in Fig.\s\ref{f:Cu-Sn}(b). When heated again (orange curve) the melting of Sn is clearly enhanced respect to the first cycle, and this time the melting is also clearly observed in the thermodiffractograms from ramp 5, as an increase in the background. This is illustrated in the projection shown in Fig.\s\ref{f:S4-2nd-heating}, in particular in the right-upper corner.

The dwell at 300$^{\circ}$C is labelled as stage 6. According to the diagram in Fig.\s\ref{f:Cu-Sn}(b), we may now expect a fraction of the sample still in the $\varepsilon$ phase with a fixed Cu$_3$Sn composition (left-side of the diagram), and the remaining fraction with a composition around 82 at.\% Sn evolving towards an $\eta + L$ state, where $L$ is again a Sn-rich liquid. The thermodiffractograms are compatible with this picture, showing the coexistence of $\varepsilon$ and $\eta$ reflections (Fig.\s\ref{f:heating1vs2}(b)). However, the relative intensity of the reflection at 63$^{\circ}$ results too high respect to the Cu$_3$Sn crystallographic model. One possible explanation for this may be in the loss of randomness in the orientation of crystals in the neutron beam. The as--quenched sample had been ground in a mortar to obtain a powder, but the subsequent transformations inside the sample holder can very well lead to a non-uniform distribution of crystal orientations, resulting in intensity mismatches respect to a perfect powder. 

Further heating to 450$^{\circ}$C (ramp 7) shows the transition from $\eta$ to $\varepsilon$, and the stabilization of $\varepsilon$ during the final dwell at 450$^{\circ}$C (labelled 8) as highlighted in Fig.\s\ref{f:sequence450}(b). Once more, the DSC curves support this qualitative image, showing during the second and subsequent cycles a much greater amount of Sn-rich melting and solidification around 200$^{\circ}$C (orange curve in Fig.\s\ref{f:S4-dsc}), and a less intense peak for the $\eta \leftrightarrow \varepsilon$ transformation at 460$^{\circ}$C. 

When turning to samples S5 and S6, with 12 at.\% In and 20 at.\% In, respectively, we arrive at a much simpler scenario, given that the thermal treatment applied \textit{in situ} at ILL was not enough for the observation of the $\eta \rightarrow \varepsilon$ transformation. For both samples, during ramp 2 there are no changes. The dwell 3 at 450$^{\circ}$C shows no further modifications, and neither does the second heating ramps 5 and 7. This is a case of a reversible process with no changes in the constitution and composition of the $\eta$ phase. DSC curves collected between 20$^{\circ}$C and 500$^{\circ}$C show neither the low--temperature peaks associated to LT-$\eta \rightarrow$ HT-$\eta$ and melting of tin, nor the high--temperature $\eta \rightarrow \varepsilon$ transition.

\section{Conclusions}

The experimental study of the phase diagram and associated phase transitions in the ternary Cu-In-Sn has become of fundamental technological relevance, given the scarce information available in the literature. \textit{In situ} neutron thermodiffraction has been used for the first time in this system. The present work gives clear evidence that precise and valuable information can be obtained when this technique is applied carefully, in combination with high-resolution diffraction to correctly identify the phases involved, and with calorimetric experiments to correlate transformation temperatures. Even the most subtle features observed in the thermodiffractograms could be successfully interpreted in terms of the possible phase equilibria proposed in the assessment by Velikanova \textit{et al.}\s\cite{07Vel}. We have shown that the metastable as--quenched state consists mostly of HT-$\eta$ phase but also some small amount of $\beta-$Sn, following the sequence of transformations $\eta \rightarrow (\eta + L) \rightarrow (\varepsilon + L)$ with transformation temperatures of 210$^{\circ}$C and 445$^{\circ}$C, respectively. The second heating cycle allowed to further explore the phase diagram after a decomposition at 450$^{\circ}$C. Our results confirm that a 5 at.\% In solubility is possible in the structure of the HT-$\eta$ phase, with no In seggregation as no unexplained reflections were found in the present study. Samples with 12 and 20 at.\% In were found to remain in a stable single-phase $\eta$ structure up to 500$^{\circ}$C. Perhaps the most important outcome of the present work is showing that the neutron thermodiffraction technique is a powerful tool to explore this system, allowing precise \textit{in situ} measurements. We believe that this will become a valuable tool in the current process of evaluation of Pb-free solders' phase diagrams\s\cite{08Din}.

\ack

This work is part of a research project supported by Agencia Nacional de Promoci\'on Cient\'ifica y Tecnol\'ogica, under grant PICT2006-1947. We particularly acknowledge ILL and the spanish CRG staff at D1B for the beamtime allocation and technical assistance. 

%\clearpage

%%%%%%%%%%%%%%%%%%%%%%%%%%%%%%%%%%%%%%%%%%%%%%%%%%%%%%%%%%%%%%%%%%%%%%%%%%%%%%%%%%

\section*{References}
\bibliography{/home/gaby/Documents/Cu-Sn-In/biblio/Cu-Sn-In.bib,/home/gaby/Documents/CV/papers-Aurelio-al-2010.bib}

\begin{thebibliography}{10}

\bibitem{08Din}
A~T Dinsdale, A Kroupa, J V{\'i}zdal, J Vrestal, A Watson, A Zemanova,
  \emph{COST Action 531-Atlas of Phase Diagrams for Lead-free Solders} { \bf
  1}, 182 (2008).

\bibitem{05Lau}
T Laurila, V Vuorinen, J Kivilahti, \emph{Materials Science and Engineering: R:
  Reports} { \bf 49}, 1 (2005).

\bibitem{07Vel}
T Velikanova, M Turchanin, O Fabrichnaya.
\newblock \emph{Cu-In-Sn (Copper-Indium-Tin)}, page 249.
\newblock Non-Ferrous Metal Ternary Systems. Selected Soldering and Brazing
  Systems: Phase Diagrams, Crystallographic and Thermodynamic Data - New Series
  IV/11C3. Materials Science International Team MSIT®, 70507 Stuttgart,
  Germany (2007).

\bibitem{72Koe}
W Koester, T Goedecke, D Heine, \emph{Zeitschrift fuer Metallkunde} { \bf 63},
  802 (1972).

\bibitem{08Lin}
S~K Lin, C~F Yang, S~H Wu, S~W Chen, \emph{Journal of Electronic Materials} {
  \bf 37}, 498 (2008).
\newblock 10.1007/s11664-008-0380-0.

\bibitem{09Lin}
S~K Lin, T~Y Chung, S~W Chen, C horng Chang, \emph{Journal of Materials
  Research} { \bf 24}, 2628 (2009).

\bibitem{90Mas}
T~B Massalski, \emph{Binary Alloy Phase Diagrams 2nd. Edition} { \bf } (1990).

\bibitem{10Nog}
K Nogita, \emph{Intermetallics} { \bf 18}, 145 (2010).

\bibitem{10Sch}
U Schwingenschl{\"o}gl, C~D Paola, K Nogita, C~M Gourlay, \emph{Applied Physics
  Letters} { \bf 96}, 061908 (2010).

\bibitem{03Aur}
G Aurelio, A~F Guillermet, G~J Cuello, J Campo, \emph{Metallurgical and
  Materials Transactions A} { \bf 34}, 2771 (2003).

\bibitem{05Aur-a}
G Aurelio, A {Fern\'{a}ndez Guillermet}, G~J Cuello, J Campo, \emph{Journal of
  Nuclear Materials} { \bf 341}, 1 (2005).

\bibitem{06Mar}
J Mart\'{\i}nez, G Aurelio, G~J Cuello, S~M Cotes, J Desimoni, \emph{Materials
  Science and Engineering: A} { \bf 437}, 323 (2006).

\bibitem{12Aur-a}
G {Aurelio}, S~A {Sommadossi}, G~J {Cuello}, \emph{ArXiv e-prints} { \bf }
  (2012).

\bibitem{LAMP}
{LAMP, the Large Array Manipulation Program}.

\bibitem{94Lar}
A~K Larsson, L Stenberg, S Lidin, \emph{Acta Crystallogr.B} { \bf 50}, 636
  (1994).

\bibitem{09Nog}
K Nogita, C Gourlay, T Nishimura, \emph{JOM Journal of the Minerals, Metals and
  Materials Society} { \bf 61}, 45 (2009).
\newblock 10.1007/s11837-009-0087-6.

\end{thebibliography}
\bibliographystyle{personal-style}

\end{document}